%% file: CKM2010_guadagnoli_7pages.tex
\documentclass[12pt]{article}
\pdfoutput=1
\usepackage{graphicx}

\newcommand\pubnumber{LPT Orsay 10-95}
\newcommand\pubdate{\today}

\def\addr{Laboratoire de Physique Th\'eorique, Universite Paris-Sud, 
Centre d'Orsay, F-91405 Orsay-Cedex, France}
\def\support{\footnote{Work supported by the Excellence Cluster `Universe', 
TUM, Munich, Germany, and by the EU Marie Curie IEF Grant. no. 
PIEF-GA-2009-251871.}}

\def\Title#1{\begin{center} {\Large #1 } \end{center}}
\def\Author#1{\begin{center}{ \sc #1} \end{center}}
\def\Address#1{\begin{center}{ \it #1} \end{center}}

\newcommand\pubblock{\rightline{\begin{tabular}{l} \pubnumber\\
         \pubdate  \end{tabular}}}
\newenvironment{Abstract}{\begin{quotation}  }{\end{quotation}}
\newenvironment{Presented}{\begin{quotation} \begin{center} 
             PRESENTED AT\end{center}\bigskip 
      \begin{center}\begin{large}}{\end{large}\end{center} \end{quotation}}
\def\Acknowledgements{\bigskip  \bigskip \begin{center} \begin{large}
             \bf ACKNOWLEDGEMENTS \end{large}\end{center}}
\usepackage{amssymb}
\newcommand{\<}{\langle}
\renewcommand{\>}{\rangle}
\newcommand{\beq}{\beqn}
\newcommand{\eeq}{\eeqn}
\newcommand{\beqn}{\begin{eqnarray}}
\newcommand{\eeqn}{\end{eqnarray}}

\newcommand{\ov}{\overline}

\newcommand{\eps}{\epsilon}

\newcommand{\rhob}{\bar \rho}
\newcommand{\etab}{\bar \eta}

\newcommand{\keps}{\kappa_{\epsilon}}

\newcommand{\mc}{\mathcal}
\newcommand{\epe}{\epsilon^\prime/\epsilon}

\newcommand{\re}{{\rm Re}}
\newcommand{\im}{{\rm Im}}
\input econfmacros.tex
\begin{document}
\begin{titlepage}
\pubblock

\vfill
\Title{On the consistency between CP violation in the $K^0$ vs. $B^0_d$ systems 
within the Standard Model}
\vfill
\Author{Diego Guadagnoli\support}
\Address{\addr}
\vfill
\begin{Abstract}
\noindent In the $K^0$ and $B^0_d$ systems, indirect CP violation is quantified by the parameters
$\eps_K$ and $\sin 2 \beta$ respectively. Within the Standard Model, the uniqueness
of the CP violating phase implies that the measurement of either between $\eps_K$ and
$\sin 2 \beta$ permits to predict the other. Since both these parameters are very well
measured, this turns into a powerful test of consistency. I discuss the status of this
test, especially in the light of recent advances on the $\eps_K$ formula.
\end{Abstract}
\vfill
\begin{Presented}
CKM2010, the 6th International Workshop on the CKM Unitarity Triangle, 
University of Warwick, UK, 6-10 September 2010
\end{Presented}
\vfill
\begin{center}
{\em In memory of Prof. Nicola Cabibbo}
\end{center}
\vfill
\end{titlepage}
\def\thefootnote{\fnsymbol{footnote}}
\setcounter{footnote}{0}
%

% SCHEME
\section{\boldmath Short statement of the problem}\label{sec:intro}

Within the Standard Model (SM), all of the wide phenomenology of flavor and CP 
violating (CPV) transitions is described in terms of one unitary matrix.
Taking into account unphysical phase redefinitions in the quark fields, the CKM 
matrix \cite{CKM}, $V$, can be parameterized in terms of 3 Euler angles and one phase $\delta$. 
Given the simplicity of this picture, the known high sensitivity of flavor and CPV 
processes to physics beyond the SM, and the amount of data at our disposal, the 
verification of the CKM mechanism is one of the most important SM tests.

In this contribution we focus on CP violation, in particular on its realization in
meson-antimeson mixings, so-called `indirect'. The latter is very accurately known
in the $K^0 - \ov K^0$ system, from the parameter $\eps_K$, of O($10^{-3}$), and in
the $B^0_d - {\ov B}^0_d$ system, from the parameter $\sin 2 \beta$, of O(1) instead.
Data are now available also for the $B^0_s - {\ov B}^0_s$ system, but, while very 
interesting, they are not yet comparable in accuracy with those on $\eps_K$ and $\sin 2 \beta$, 
and I will confine my discussion to the latter two quantities. Because of the unique CPV 
phase $\delta$, the parameters $\eps_K$ and $\sin 2 \beta$ are necessarily 
correlated within the SM, and this offers one of the most stringent tests of the 
CKM picture of CPV.

The main conclusion of this contribution is that the performance of the SM in the $\eps_K -
\sin 2 \beta$ correlation is less than perfect. In short, the problem can be stated
as follows. Since the dominant contribution to $\eps_K$ is proportional to 
$\im (V_{ts} V^*_{td})^2$, with $V_{ts} \simeq - A \lambda^2$ and $V_{td} \simeq 
A \lambda^3 (1-\rhob-i\etab)$ \cite{Wolf}, one has $\eps_K \propto (1-\rhob) \etab$. 
Then, recalling that $1-\rhob = R_t \cos \beta$ and $\etab = R_t \sin \beta$ (with 
$R_t$ and $\beta$ one of the sides and one of the angles of the `unitarity triangle' 
implied by the CKM matrix \cite{Buras_LH}), one sees that $\eps_K \propto \sin 2 \beta$. 
Because the proportionality factor is calculable, knowledge of either between $\eps_K$ and 
$\sin 2 \beta$ allows to predict the other.
Concerning this proportionality factor, it has been pointed out in recent work \cite{BG1,BG2,BGI}
that subleading contributions to the $\eps_K$ formula -- the main topic of this contribution -- 
can be bundled in a non-negligible, negative, multiplicative correction, $\keps$, to 
$\eps_K$. Therefore, for fixed $\sin 2 \beta$, this correction would bring the $|\eps_K|$
central value down. Numerically, if one plugs in the golden-mode determination 
$\sin 2 \beta_{\psi K_S} \simeq 0.67$, one finds a central value for $|\eps_K|$ of about 
$1.85 \times 10^{-3}$, to be compared with $|\eps_K|^{\rm exp} \simeq 2.2 \times 10^{-3}$.%
\footnote{Equivalently, enforcing the experimental $|\eps_K|$ constraint instead, one finds 
a prediction for $\sin 2 \beta$ higher than the experimental determination \cite{LS}. 
% Interestingly, it has been recently pointed out that the most recent data on 
% the BR$(B \to \tau \nu)$ \cite{Btaunu_exp} also hint at $\sin 2 \beta > \sin 2 \beta_{\psi K_S}$ 
% \cite{LS_Btaunu}.
}
Without (yet) any reference to associated errors, this difference seems to indicate
that the level of consistency of the CKM picture for CPV is at no better than the 20\%
level (!). These statements will be made more quantitative in section \ref{sec:error}.

\section{\boldmath The $\eps_K$ formula and the $\keps$ correction}\label{sec:epsK}

A general theoretical formula for $\eps_K$ is the following
\begin{equation}
\label{eq:epsK_BG}
\eps_K ~=~ e^{i \phi_\eps} \sin \phi_\eps \left( \frac{\im M_{12}}{\Delta M_K} +\xi \right)~,
\end{equation}
where $\Delta M_K \equiv m_{K_L} - m_{K_S} \simeq 3.5 \times 10^{-15}$ GeV and 
$\Delta \Gamma_K \equiv \Gamma_{K_L} - \Gamma_{K_S} \simeq - 7.4 \times 10^{-15}$ GeV. 
Accidentally, $\Delta \Gamma_K \simeq -2 \Delta M_K$, hence it is useful to define the phase 
$\phi_\eps \equiv $ $\arctan (-\Delta M_K/\frac{1}{2}\Delta \Gamma_K) \simeq 43.5^\circ$.
The other quantities in eq. (\ref{eq:epsK_BG}) are the amplitude for $\ov K^0 - K^0$ mixing,
$M_{12} = \< K^0 | \mc H_{\Delta S = 2} | \ov K^0 \>$, that is the part sensitive to non-SM 
contributions, and $\xi$, quantity on which I will return in more detail below and in section 
\ref{sec:OPE}. Typical approximations
in the literature consist in setting $\phi_\eps = 45^\circ$ and $\xi = 0$. Refs. \cite{BG1,BG2}
pointed out that both these corrections are negative and sum up to a total $-8\%$ correction.
In fact, the combined effect of $\phi_\eps \neq 45^\circ$ and $\xi \neq 0$ can be described 
by a multiplicative factor $\keps$ such that 
\begin{equation}
\label{eq:kepsdef}
\eps_K = \keps \, \eps_K (\phi_\eps = 45^\circ,\xi = 0)~,
\end{equation}
and one gets the estimate $\keps = 0.92 \pm 0.02$ \cite{BG1}. This estimate can be
understood as follows. By the $\keps$ definition one easily obtains
\begin{equation}
\label{eq:keps}
\keps ~=~ \frac{\sin \phi_\eps}{1/\sqrt 2} \times \left( 1 + \frac{\xi}{\sqrt 2 |\eps_K|} \right)~,
\end{equation}
where, in view of the smallness of $\xi$, we identify $|\eps_K|$ on the r.h.s. with $|\eps_K(\phi_\eps = 45^\circ,\xi = 0)|$.
The hard part is a quantitative estimate of $\xi$. One can however note that the combination $\xi / {\sqrt 2 |\eps_K|}$ also 
appears in the theoretical formula for the parameter $\eps'/\eps_K$, namely $|\eps'/\eps_K| = 
- \omega \frac{\xi}{\sqrt 2 |\eps_K|} (1 - \Omega)$ \cite{Buras_LH}, where $\omega = 0.045$ is known
very precisely, and quantifies the `$\Delta I = 1/2$ rule'. Enforcing equality of this formula 
with $|\eps'/\eps_K|_{\rm exp} =(1.65\pm0.26) \cdot 10^{-3}$, the problem of estimating $\xi$ is restated 
into that of estimating $\Omega$. 
This quantity is the ratio between so-called EW-penguin and QCD-penguin contributions to $\eps' / \eps_K$:
both of them are calculable -- with caveats on the knowledge of the hadronic matrix elements -- 
using the work of refs. \cite{BJL_epe}. Proceeding as 
explained in sec. 4 of ref. \cite{BG2}, one finds $\Omega = 0.33 (1 \pm 20\%)$, assuming the SM. 
Note that the still quite poor theoretical control on, especially, the QCD penguins is reflected in the 20\%
error on $\Omega$, and the $\Omega$ estimate, in turn, results in $\xi/\sqrt2 |\eps_K| = -0.054 (1
\pm 25\%)$, where the error takes into account also the $15\%$ experimental error on $\epe_K$.
It is to be observed that the large final error, 25\%, on $\keps$, affects only marginally the total 
error on $\eps_K$, since it is an error on a correction.

\section{\boldmath A closer look at the OPE for $\eps_K$}\label{sec:OPE}

\begin{figure}[t]
\centering
\includegraphics[width=0.35\textwidth]{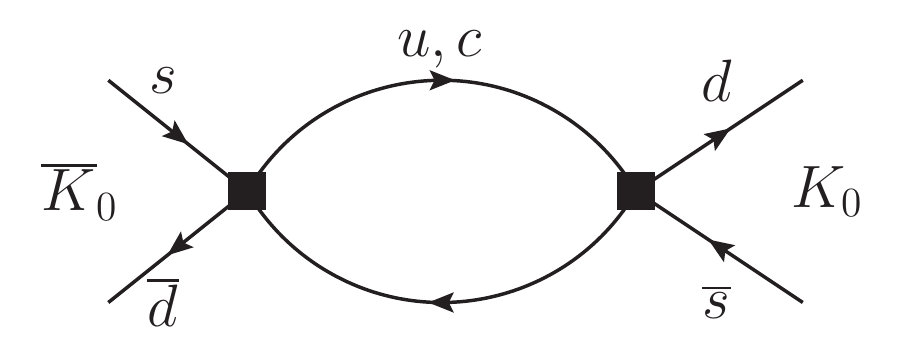}
\includegraphics[width=0.35\textwidth]{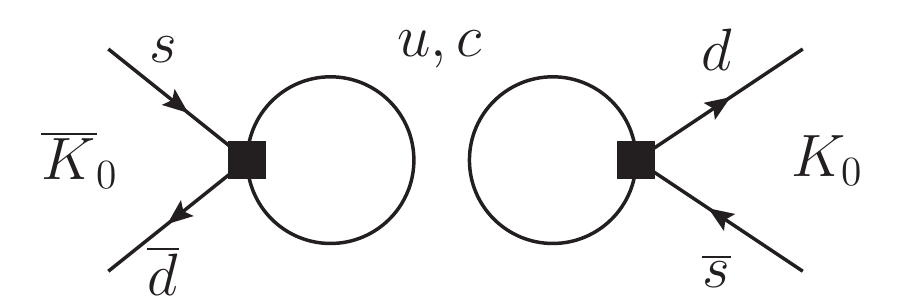}
\caption{Contractions of the leading $|\Delta S| = 1$ four-quark effective operators contributing
to the mixing amplitude at O($G_F^2$).}
\label{fig:diags}
\end{figure}

Let us now look more closely at the meaning of the parameter $\xi$. This parameter is defined as
$\xi = \im A_0 / \re A_0$, with $A_0$ the amplitude for $K^0 \to (\pi \pi)_{I=0}$. A more inspiring 
interpretation arises by observing that the $|(\pi\pi)_{I=0}\>$ state largely saturates the neutral
kaon decay widths. Denoting the absorptive part of the $\ov K^0 - K^0$ amplitude as $\Gamma_{12}$, 
one has therefore
\begin{equation}
\label{eq:Gamma12}
\Gamma_{21} = \Gamma_{12}^* = \sum_{f} \mc A(K^0\to f) \mc A(\bar K^0 \to f)^* \simeq (A_0)^2~,
\end{equation}
where, as mentioned, in the last step we have approximated the final states to consist only
of $|(\pi\pi)_{I=0}\>$. Eq. (\ref{eq:Gamma12}) implies
\begin{equation}
\label{eq:xi}
\frac{\im \Gamma_{12} }{ \re\Gamma_{12} } \approx - 2 \frac{\im A_0}{\re A_0} = - 2 \xi~.
\end{equation}
Therefore $\xi$ is generated by the imaginary part of the {\em absorptive} contribution to the mixing
amplitude, namely, by the absorptive part of the diagrams in Fig. \ref{fig:diags}.
Comparing $\xi$ as in eq. (\ref{eq:xi}) with the $\eps_K$ formula (\ref{eq:epsK_BG}), one realizes \cite{BGI}
that, if one is to include $\xi$, one should also compute and include for consistency the {\em dispersive} 
part of the diagrams in Fig. \ref{fig:diags}, because $\re M_{12}$ and $\re \Gamma_{12}$, appearing in the 
denominators (recall that $\Delta M_K \simeq 2 \re M_{12}$) are of the same order.

A consistent, beyond-leading-order evaluation of $\eps_K$ requires therefore to take into account:
{\em (a)} non-local contributions to both $\im M_{12}$ and $\im \Gamma_{12}$ generated by the O$(G_F)$ 
dimension-six $\Delta S=1$ operators; {\em (b)} local contributions to $\im M_{12}$ generated by 
dimension-eight $\Delta S=2$ operators of O$(G_F^2)$. The non-local contributions to $\im \Gamma_{12}$,
giving rise to $\xi$, have been estimated in Ref. \cite{BG1} with the strategy outlined beneath
eq. (\ref{eq:keps}). 

Concerning $\im M_{12}$, it assumes the form $\im  M_{12} = \im M_{12}^{SD} + \im M_{12}^{LD}$, 
with $\im M_{12}^{LD} = \im M_{12}^{\rm non-local} + \im M_{12}^{(8)}$,
where $\im M_{12}^{\rm non-local}$ and $\im M_{12}^{(8)}$ are not separately scale independent. 
The structure of the dimension-eight operators obtained integrating out the charm, and an estimate of 
their impact on $\eps_K$, has been presented in Ref.~\cite{CataPeris}. According to this estimate, 
$\im M_{12}^{(8)}$ is less than $1\%$ of the leading term.

The only potentially large long-distance contribution to $\im M_{12}$ is therefore the contribution of 
the non-local terms enhanced by the $\Delta I=1/2$ rule. Given their long-distance nature, these terms
can be described within Chiral Perturbation Theory (ChPT), the limit of QCD valid at scales below the 
threshold for production of vector-meson states. Within ChPT, the $\Delta I=1/2$ part of $\mc H_{\Delta S=1}$ 
consists of one operator only, with effective coupling $G_8$. Then by definition one must have $A_0 \propto G_8$.
Thus, decomposing $\im M_{12}^{LD}$ as a leading term proportional to $G_8^2$, plus a subleading term 
with different effective coupling, namely $\im M_{12}^{LD} = \im M_{12}^{LD}|_{G_8^2} + 
\im M_{12}^{LD}|_{{\rm non}-G^2_8}$, one can write 
\begin{equation}
\left.\im M_{12}^{LD}\right|_{G_8^2} = \left.\re M_{12}^{LD}\right|_{G_8^2} 
\times  \frac{\im [(G_8^*)^2] }{ \re [(G_8^*)^2] }~,
\end{equation}
whence, using $A_0 \propto G_8$ as mentioned, and eq. (\ref{eq:xi}) one arrives at
\begin{equation}
\left.\im M_{12}^{LD}\right|_{G_8^2} \approx  \left.\re M_{12}^{LD}\right|_{G_8^2} 
\times (-2\xi) \approx  - \xi \times \left(
\Delta m^{LD}_K|_{G_8^2} \right)~.
\label{eq:main}
\end{equation}
This allow us to rewrite eq. (\ref{eq:epsK_BG}) as follows \cite{BGI}
\begin{equation}
\label{eq:epsK_BGI}
|\eps_K|  = \sin \phi_\eps \Bigl[ \frac{\im M^{(6)}_{12}}{\Delta m_K } + 
\xi \underbrace{\left( 1 -\frac{\Delta m^{LD}_K|_{G_8^2}}{\Delta m_K} \right)}_{\rho} + \delta_{ \im M_{12} } \Bigl]~,
\end{equation}
where $\delta_{ \im M_{12} }$ encodes any contribution to $\im M_{12}^{LD}$ {\em not} generated
by the double insertion of the $G_8$ operator, and also $\im M_{12}^{(8)}$. Eq. (\ref{eq:epsK_BGI})
defines the parameter $\rho$, whose deviation from 1 quantifies long-distance effects to $\im M_{12}$.
The estimate of the parameter $\rho$ within ChPT is summarized in the section to follow.

\subsection{\boldmath Estimate of long-distance effects to $\im M_{12}$ within ChPT} \label{sec:ChPT}

The lowest-order ChPT Lagrangian describing non-leptonic $\Delta S=1$ decays has only two operators,
transforming as $(8_L,1_R)$  and $(27_L,1_R)$ under the  $SU(3)_L\times SU(3)_R$ chiral group \cite{CI}. 
Of these operators, only the $(8_L,1_R)$ one has a phenomenologically large coefficient, describing 
the observed enhancement of $\Delta I=1/2$ amplitudes. Hence, the only term in the effective Lagrangian 
relevant to our calculation is
\begin{equation}
\label{eq:ChPTDS=1}
\mc L^{(2)}_{|\Delta S| = 1} = F^4  G_8 \left( \partial^\mu U^\dagger \partial_\mu U \right)_{23} +{\rm h.c.} ~,
\end{equation}
where, as usual, the set of pion fields $\Phi$ is defined in $U =\exp(i \sqrt{2} \Phi/F)$ and $F$ can be 
identified with the pion decay constant ($F\approx 92\, {\rm MeV}$). The fact that the $(8_L,1_R)$ operator
must describe $\Delta I=1/2$ amplitudes implies that, at tree level, the phase of $G_8$ coincides with $\xi$:
$\xi \equiv \im A_0/ \re A_0 = \im(G_8)/\re(G_8)$. 
Furthermore, from the full ChPT formula for $A_0$ one can also estimate the $G_8$ magnitude to be 
O($G_F$), $|G_8|\approx 9\times 10^{-6}~({\rm GeV})^{-2}$ \cite{CI}. Hence the full $G_8$ coupling can be 
determined from data on $K\to 2\pi$ amplitudes.

\begin{figure}[t]
\begin{center}
\raisebox{0.86cm}{\includegraphics[width=0.35\textwidth]{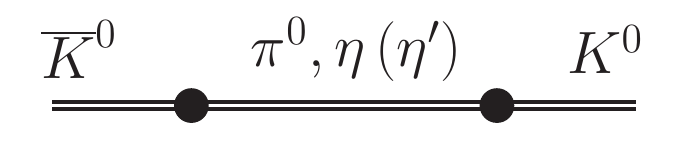}}
\hspace{1cm}
\includegraphics[width=0.35\textwidth]{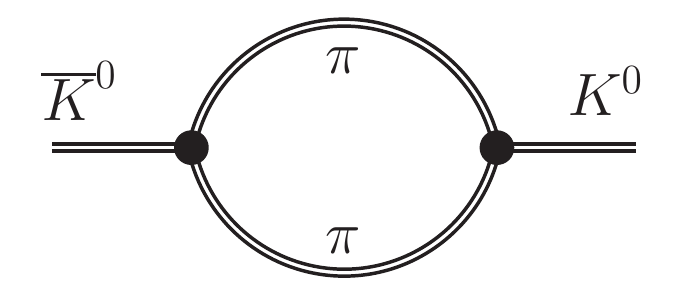}
\end{center}
\caption{\label{fig:diagsChPT} Tree-level and one-loop diagrams contributing to $\bar K^0$--$K^0$ 
mixing within ChPT.}
\end{figure}

The ChPT diagrams contributing to $\Delta m_K$ up to O$(p^4)$ are depicted in Fig. \ref{fig:diagsChPT}.
The tree-level, O$(p^2)$, diagram in the left panel vanishes when using the so-called 
Gell-Mann--Okubo formula, namely the O$(p^2)$ relation among $\pi^0$, $\eta$ and kaon masses 
\cite{Donoghue:M12}.
As a result, the first non-vanishing contribution to $M_{12}$ generated by 
$\mc L^{(2)}_{|\Delta S| = 1}$ arises only at O$(p^4)$.

At O$(p^4)$ the amplitude to be calculated consists of loop contributions with two insertions 
of $\mc L^{(2)}_{|\Delta S| = 1}$, plus tree-level contributions with the insertion 
of appropriate O$(p^4)$ counterterms, cancelling the renormalization scale dependence.
Among all these O$(p^4)$ contributions, the only model-independent, and presumably dominant, 
contribution to $M_{12}$ is the non-analytic one generated by the pion-loop amplitude, 
$A^{(\pi \pi)}$, in 
Fig.~\ref{fig:diagsChPT} (right). Our calculation has been confined to this contribution, for 
the reasons to follow: (1) $A^{(\pi \pi)}$ is the only contribution which has an absorptive 
part. Hence it's the only contribution whose weak phase (the information relevant to our calculation)
can be extracted from data; 
(2) It is the only contribution that survives in the limit of $SU(2)_L \times SU(2)_R$ ChPT, 
which is known to be successful phenomenologically (see e.g.~Ref.~\cite{CI});
(3) In the words of ref. \cite{Donoghue:SU3ChPT}, it is almost a theorem that kaon loops go 
into the redefinition of the local terms; and besides, there are doubts about the meaning of 
kaon loops, their effective threshold lying at $2 m_K > m_\rho$ (!) \cite{Donoghue:SU3ChPT}.

\subsection{\boldmath Final phenomenological formula for $\eps_K$}

Using the calculation of $A^{(\pi \pi)}$ described in the previous section (for more details, see
ref. \cite{BGI}), we can estimate the contribution to  $\im M_{12}$ proportional to $G_8$, 
namely the $\rho$ parameter entering eq. (\ref{eq:epsK_BGI}). We find \cite{BGI}
\begin{equation}
\frac{\Delta m^{LD}_K|_{G_8^2}}{\Delta m_K^{\rm exp}} = \frac{2 \re M_{12}^{\rm (\pi \pi)} }{\Delta m_K^{\rm exp}} = 0.4 \pm 0.2~.
\label{eq:chpt_main}
\end{equation}
where the central value and error are determined by setting the renormalization scale $\mu =800$~MeV 
and respectively varying it in the interval $0.6 \div 1$~GeV.

As a cross-check, our calculation shows good agrement with what one would expect from the 
rest of the contributions to $\Delta m_K$ known as dominant. One has roughly 
$\rho \simeq (\Delta m^{SD}_K + \Delta m^{LD}_K|_{\eta^\prime})/\Delta m_K^{\rm exp}$,
with $\Delta m_K^{(6)} = (0.7\pm 0.1) \Delta m_K^{\rm exp}$ \cite{DMk_SD}
and $\Delta m^{LD}_K|_{\eta^\prime} \approx -0.3 \Delta m_K^{\rm exp}$ according to the
comprehensive analysis of the $\eta^\prime$ exchange amplitude in Ref. \cite{GST}.

{\bf To recapitulate,} our final phenomenological formula for $\eps_K$ is eq. (\ref{eq:epsK_BGI}), with
$\rho = 0.6 \pm 0.3$ \cite{BGI} (we conservatively increase the error by 50\% to account for the subleading
non-$G_8^2$ contributions). In terms of $\keps$ defined in eq. (\ref{eq:kepsdef}), this implies
$\keps = 0.94 \pm 0.02$.

\section{\boldmath Error budget in $\eps_K$, and status of the problem} \label{sec:error}

In sec. \ref{sec:intro} we argued that the $|\eps_K^{\rm SM}|$ central value is 
about 20\% beneath the experimental figure. Of course the significance of this difference
is in the hands of the associated errors. The main components to the $|\eps_K^{\rm SM}|$ error
can be understood intuitively by focusing on the top-top contribution to $\eps_K^{\rm SM}$, 
which constitutes about 75\% of the total result. They are as follows. 
First, the calculation of $\im M_{12}^{(6)}$ involves 
estimating the non-perturbative matrix element between the single dim-6 SM operator
and the external kaon states. This affects linearly $|\eps_K|$, via the parameter $\hat B_K$, 
which is known with an uncertainty of O(5\%). For recent unquenched LQCD estimates, see 
Refs. \cite{BK_recent}. For an upper limit on $\hat B_K$ using large $N_c$, I also
signal Ref. \cite{Gerard_BKbound}.

Second, there is the CKM error. One way to parameterize the CKM matrix useful in this discussion
is through the parameters $\lambda$, $\sin 2 \beta$, $|V_{cb}|$ and $R_t$. The first
two parameters are very well known, so let's focus on the other two. Recalling the CKM
combination relevant to $|\eps_K^{\rm SM}|$, namely $(V_{ts} V^*_{td})^2$, and that 
$|V_{ts}| \approx |V_{cb}|$ and $|V_{td}| \approx \lambda |V_{cb}|$, one sees that 
$|\eps_K^{\rm SM}| \propto |V_{cb}|^4$. A $|V_{cb}|$ error presently at the 2.5\% level
translates into a 10\% error on $|\eps_K^{\rm SM}|$. This is the dominant source of error. 
Finally, recalling from sec. \ref{sec:intro} that $\eps_K \propto (1-\rhob) \etab$, 
one finds $\eps_K \propto R_t^2$, contributing another O(8\%) component
to the $\eps_K$ relative error. Fortunately, this component is going to become irrelevant 
once a very accurate determination of the CKM angle $\gamma$ will be available from LHCb
data. The above mentioned three components build up a total error of order 15\% in 
$|\eps_K^{\rm SM}|$.

I open a parenthesis on two further directions relevant to a more precise assessment of the 
$\eps_K - \sin 2 \beta$ SM correlation. On the one side, the effort towards a NNLO calculation 
of the coefficient $\eta_{ct}$
\cite{BrodGorbahn}, appearing in the short-distance calculation. Note in fact that there are
cancellations between the $t$-$t$ (73\%), $c$-$t$ (41\%) and $c$-$c$ ($-$14\%) contributions, so
each of them should be known with the best possible accuracy. On the other side, the possible
role of doubly-Cabibbo-suppressed corrections to the CP asymmetries leading to the $\sin 2 \beta$
estimate \cite{penguins}.

To conclude, the most rigorous way to quantify a possible problem in the $\eps_K - \sin 2 \beta$ SM 
correlation remains that of a fit \cite{CKM_pages}. The UTfit group confirms that the new 
contributions in $\eps_K$ generate some tension in particular between $\eps_K$ and $\sin 2 \beta$, 
tension manifested by an indirect determination of $\sin 2 \beta$ larger than the experimental 
value by about 2.6$\sigma$ (see \cite{UTfit_ICHEP2010}). Furthermore, the overall picture is 
very usefully summarized in so-called compatibility plots. 
The CKMfitter group, with the `Rfit' treatment of theory errors, finds no 
discrepancy \cite{LN_CKMfitter}. This treatment may however be too conservative, 
and in fact they conclude that the matter needs further investigation.
 
\Acknowledgements

\noindent I am thankful to Andrzej J. Buras and Gino Isidori for the most pleasant collaboration.
I also thank the organizers of WG1 at CKM2010, in particular Federico Mescia, for the kind invitation
to this stimulating workshop.

\end{document}

%% file: econfmacros.tex
\def\beq{\begin{equation}}
\def\eeq#1{\label{#1}\end{equation}}
\def\eeqn{\end{equation}}

\def\beqa{\begin{eqnarray}}
\def\eeqa#1{\label{#1}\end{eqnarray}}
\def\eeqan{\end{eqnarray}}

\let\bar=\overbar

\def\Dslash{\not{\hbox{\kern-4pt $D$}}}
\def\dslash{\not{\hbox{\kern-2pt $\del$}}}

\def\msb{{\bar{\ssstyle M \kern -1pt S}}}

\def\eps{\epsilon}